\documentclass[conference]{IEEEtran}
\usepackage{cite}
\usepackage{amsmath,amssymb,amsfonts}
\usepackage{graphicx}
\usepackage{textcomp}
\usepackage{xcolor}
\usepackage{braket}
\usepackage{caption}
\usepackage{subcaption}
\usepackage{algorithm}
\usepackage[noend]{algpseudocode}
\usepackage{array}
\usepackage{enumitem}
\usepackage{tikz}

\def\BibTeX{{\rm B\kern-.05em{\sc i\kern-.025em b}\kern-.08em
    T\kern-.1667em\lower.7ex\hbox{E}\kern-.125emX}}
\begin{document}

\algnewcommand{\Input}[1]{\State \textbf{input} #1}
\algnewcommand{\Output}[1]{\State \textbf{output} #1}
\algrenewcommand{\Return}[1]{\State \textbf{return} #1}
\algnewcommand{\LineComment}[1]{\State \(\triangleright\) #1}

\newcolumntype{V}[1]{>{\vspace*{0.25em}\centering\arraybackslash} m{#1} <{\vspace*{0.25em}}}

\title{Need One Bell-pair Only (NOBOL) for Low-Overhead Fault-Tolerant Quantum Computing}

\author{\IEEEauthorblockN{1\textsuperscript{st} Sean Grzenda}
\IEEEauthorblockA{\textit{Computer Science and Engineering} \\
\textit{University at Buffalo}\\
Buffalo, USA \\
seangrze@buffalo.edu}
\and
\IEEEauthorblockN{2\textsuperscript{nd} Shahram Babaie}
\IEEEauthorblockA{\textit{Computer Science and Engineering} \\
\textit{University at Buffalo}\\
Buffalo, USA \\
shahramb@buffalo.edu}
\and
\IEEEauthorblockN{3\textsuperscript{rd} Chunming Qiao}
\IEEEauthorblockA{\textit{Computer Science and Engineering} \\
\textit{University at Buffalo}\\
Buffalo, USA \\
qiao@buffalo.edu}
}

\maketitle

\begin{abstract}
Fault-tolerant quantum computation fundamentally relies on encoding a logical qubit into a structured block of physical qubits, typically in the tens to hundreds. As a trade-off for improved fault-tolerance, logical gate operations will incur a linear overhead, in terms of both the amount of time and quantum resources, than before. For example, in monolithic quantum computing, performing a gate operation on two distant logical qubits will first require using a linear number of SWAP operations in order to move the logical qubits next to each other; while in distributed quantum computing, doing so will first require a linear number of ancilla qubits in order to form entanglement connections (or a logical Bell pair). 

In this paper, we focus on significantly reducing the overhead involved in logical CNOT operations, a fundamental primitive. We propose NOBOL, a novel approach that requires only one Bell pair to perform a logical CNOT operation on two distant qubits encoded in arbitrary Calderbank–Shor–Steane (CSS) codes. More importantly, NOBOL only requires performing gate operations on the logical X or Z operator subsets of the logical qubits. For many CSS codes, such as the surface code, these subsets are significantly smaller than the size of the code itself. In this paper, we describe various circuit realizations of NOBOL, including a depth optimal circuit with logarithmic depth in terms of the size of the logical operators. Finally, we propose effective methods to contain error propagation without incurring much additional overhead.

Since NOBOL can be effectively applied to a wide range of quantum error-correcting (QEC) codes, and in addition, is agnostic to qubit modalities and effective for various architectures, including those based on either a monolithic QPU or distributed QPUs, we believe that this work makes a significant contribution to advancing large-scale fault-tolerant quantum computing and Quantum Internet.
\end{abstract}


\section{Introduction}
Quantum computing offers the potential for significant speedups in a variety of applications, such as integer factorization, discovery of new materials, and combinatorial optimization by exploiting unique quantum phenomena such as superposition and entanglement  \cite{doi:10.1137/S0097539795293172, 10.1145/3708471, Cerezo2022, 128057}. Despite these promising advantages, current quantum hardware remains  highly error-prone and constrained in the number of available physical qubits (thus classified as noisy intermediate-scale quantum or  NISQ devices). 

To address the fragility of quantum states, quantum error correction (QEC) codes encode quantum information across a block of entangled physical qubits into one logical qubit to improve resilience to noise and errors \cite{babaie2024}. QEC enables detection and correction of errors, which forms the foundation for fault-tolerant quantum computing \cite{Campbell2024}. However, this redundancy-based mechanism further amplifies the demand for physical qubits, thereby exacerbating scalability challenges. In this context, distributed quantum computing (DQC) has emerged as a promising paradigm, where multiple small quantum processing units (QPUs) are interconnected. Through a quantum data network \cite{qiao-qdn-dqc,zhao-qdn} and classical communication links, a larger virtual QPU can be formed. DQC provides a scalable pathway to overcome qubit limitations and facilitates large-scale fault-tolerant quantum computing\cite{filippov2025architectingdistributedquantumcomputers, mohseni2026buildquantumsupercomputerscaling, 9923784, VanDamme2024, Hertzberg2021}.

Fault-tolerant quantum computing not only relies on a collection of error management techniques, including error suppression, error correction, and error containment, but also requires a fault-tolerant universal set of gate operations. A fundamental approach to constructing fault-tolerant gate implementations is through transversal operations. In CSS codes \cite{babaie2025, 11568217}, single-qubit transversal gates are realized as tensor products of independent operations applied to each physical qubit in a code block, while multi-qubit transversal gates are implemented in a bitwise manner, acting on corresponding pairs of qubits across different code blocks. 

The CNOT gate is one of the most fundamental two-qubit operations in quantum computing due to its prominent role in entanglement generation, and its contribution to forming a universal gate set (together with the $T$ and $H$ gates). It plays a central role in QEC, distributed quantum protocols, and the implementation of logical operations, which makes it a key determinant of both computational capability and system performance \cite{Mueller2025-en}. Since CNOT is a Clifford gate, for many QEC codes, CNOT has a transversal implementation, and, when the logical qubits are neighboring eachother on the same QPU, this operation is trivial. However, if the qubits are non-neighboring, then SWAP gates are required, which increases overhead. Other approaches, such as lattice surgery or lattice braiding incur similar overhead when the logical qubits are non-neighboring.

The overhead becomes worse when considering distributed systems. If the logical qubits are on separate QPUs, then several entanglement links are required to perform the operation transversely. CNOT can be implemented using parity-based-measurement approaches, such as distributed lattice surgery \cite{PhysRevA.92.042305, Huai2024, keskin2025lattice}. However, this method also requires numerous Bell pairs. It is possible to perform remote operations at the logical level using logical gate or logical qubit teleportation. Nevertheless, even these approaches incur overhead through the formation of logical Bell pairs. More specifically, this dependence on logical entanglement resources introduces substantial overhead in terms of resource consumption, communication rounds, and sensitivity to errors. In many cases, the resource cost of a distributed logical operation scales unfavorably with the code distance, further amplifying the burden of fault tolerance. Moreover, repeated entanglement generation and classical feedback introduce latency that can degrade overall system performance. In short, remote logical gate operations pose a key roadblock towards large-scale quantum computing via e.g. DQC.

In this paper, we propose NOBOL, a novel low-overhead approach for implementing logical CNOT operations that significantly reduces reliance on entangled resources and ancilla qubits. NOBOL aims to reduce the overhead in both the monolithic case, where the qubits are non-neighboring, and in the distributed case, where the qubits reside on separate QPUs. NOBOL enables the realization of a logical CNOT between two encoded qubits using only a single entangled state (e.g., a Bell pair). This Bell pair is first established and distributed at the same two locations as the two logical qubits. \emph{Thereafter, only local operations at both locations, involving one half of the Bell pair and one logical qubit at the same location, are performed}.

The proposed design eliminates the need for excessive SWAP operations when the logical qubits involved in the same CNOT operation reside on the same QPU but are non-neighboring, and greatly benefits distributed multi-QPU architectures. By requiring only a single shared entangled state, independent of the block size, NOBOL relaxes key hardware constraints associated with entanglement generation, distribution, and preservation. This feature makes it well-suited for near- and mid-term large-scale quantum platforms, where interconnects remain a critical bottleneck.

More specifically, consider a logical CNOT operation on two remote logical qubits, $A_L$ (as the control) and $B_L$ (as the target), where $A_L$ and $B_L$ are each encoded using a CSS code with block size $N$, typically in tens to hundreds. $A_L$ and $B_L$ can be encoded using any arbitrary CSS code as the method is code agnostic; it only works with the logical operators. Let $A_{Z_L}$ and $B_{X_L}$ be the subset of the physical qubits in the block encoding $A_L$ and $B_L$ respectively, that support a logical $Z$ and $X$ operators of $A_L$ and $B_L$, respectively. For simplicity, assume that the size of each subset is $\tau$, which typically can be much smaller than $N$ (for example, in a surface code, $\tau=\sqrt{N}$). Further, denote the two ancilla qubits that share the single entangled Bell state  by $q$ and $q'$, respectively. 

NOBOL implements CNOT $(A_L, B_L)$ by first establishing and distributing the Bell pair such as that $q$ and $q'$ are at the same location as $B_L$ and $A_L$, respectively to facilitate the subsequent local operations. Thereafter, it uses (1) only $2\tau-1$ \emph{local} CNOT gates on $q$ and the physical qubits in $B_{X_L}$, along with (2) an additional $2\tau-1$ \emph{local} CNOT gates on $q'$ and the physical qubits in $A_{Z_L}$. The proposed design admits a factorial number of circuit realizations due to the flexibility in qubit ordering, while its depth-optimal implementation achieves a depth of $O(\text{log} (\tau))$. This combination of structural flexibility and shallow circuit depth enables diverse mappings onto hardware architecture to accommodate hardware constraints and aligns well with coherence-time requirements in fault-tolerant quantum computing. Finally, in this paper, we introduce lightweight mechanisms to enhance fault tolerance by incorporating intermittent stabilization and encoding $q$ and $q'$ using small repetition codes. 

While this work does show it is possible to perform logical operations with a single physical Bell pair, naturally, this method will reduce the fault tolerance of the operation as redundancy has been eliminated. Thus, we provide error containment procedures for improving the fault tolerance. The issue is a single error can spread to other data qubits causing a logical error. One method to prevent the error propagation is to perform intermittent stabilization, and the other method is to encode each half of the Bell pair in small repetition codes to distribute the operation within the code block. Each error containment procedure has its own trade-off in terms of execution time and ancilla qubit use.

Given that NOBOL can be effectively applied to a wide range of quantum error-correcting (QEC) codes, including arbitrary CSS codes. and in addition, is agnostic to qubit modalities and effective for various architectures, including those based on either a monolithic QPU or distributed QPUs, NOBOL is compatible with a wide range of qubit modalities and QEC codes, we believe that this work makes a significant contribution to advancing large-scale fault-tolerant quantum computing and Quantum Internet. 

The remainder of this paper is organized as follows. Section II provides background on distributed quantum computing, logical qubits, and logical operations. Section III presents the NOBOL design, including its various structural implementations and optimal-depth constructions. Error containment techniques for fault-tolerant implementations of NOBOL are detailed in Section IV. Finally, Section V concludes the paper.

\section{Background and Preliminaries}

\subsection{Distributed Quantum Computing}

\begin{figure}
    \centering
    \includegraphics[width=0.8\linewidth]{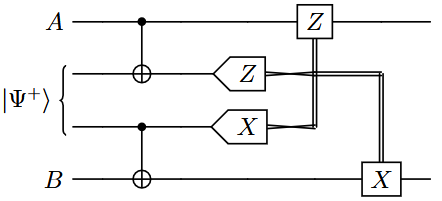}
    \caption{Remote CNOT operation between $A$ and $B$. $A$ and $B$ are physical qubits residing on two locations, e.g. QPUs. $\ket{\Phi^+}$ is a Bell pair state distributed between the two locations.}
    \label{fig:PhysRemoteCNOT}
\end{figure}

Distributed quantum computing architectures comprise multiple quantum processing units interconnected via quantum data networks (QDN) \cite{qiao-qdn-dqc,zhao-qdn} classical communication channels. Each QPU can perform local operations on its encoded logical qubits, while remote logical operations are typically enabled through the generation and consumption of entangled states between modules \cite{filippov2025architectingdistributedquantumcomputers}. This modular paradigm alleviates several limitations associated with monolithic quantum computing (MQC), including limited qubit counts, constrained qubit connectivity, control complexity, and fabrication constraints.

The dominant cost driver in DQC is the realization of remote operations across QPUs. Despite their non-local nature, such operations can be implemented using only local operations and classical communications (LOCC). The canonical example is the remote CNOT gate illustrated in Fig.~\ref{fig:PhysRemoteCNOT}, which leverages a pre-distributed Bell pair between QPUs, given by Equation \ref{eq:bell}.

\begin{equation}
\label{eq:bell}
    \ket{\Phi^+}=\frac{\ket{00}+\ket{11}}{\sqrt{2}}
\end{equation}

In this construction, all quantum gates are executed locally within each QPU, while classical communication is used to transmit measurement outcomes and apply conditional correction operations, effectively realizing a remote CNOT gate\cite{PhysRevA.62.052317}. In practice, the efficiency of such remote gate operations is fundamentally constrained by the underlying entanglement distribution techniques. Entanglement between QPUs is typically established via photonic interconnects or similar networking techniques, which are inherently probabilistic and subject to noise and loss. As a result, entanglement generation exhibits limited success probability and introduces non-negligible latency \cite{11000201, PhysRevA.107.022428}. The expected time required to establish a Bell pair of fidelity $F<1$ can be expressed by Equation \ref{eq:link}.

\begin{equation}
    T_{\text{link}} = \frac{1}{p_{\text{link}}}T_0
    \label{eq:link}
\end{equation}
where $p_{\text{link}}$ denotes the success probability of entanglement generation and $T_0$ is the duration of a single attempt \cite{doi:10.1126/science.aam9288, RevModPhys.83.33, doi:10.1126/science.aao4309}. As indicated by (\ref{eq:link}), low success probabilities directly result in increased latency, which influences remote gate operations. Consequently, the performance of remote logical gate operations is tightly coupled to the cost of entanglement generation and distribution. This dependency introduces a fundamental bottleneck in DQC systems, where communication overhead, latency, and accumulated errors from repeated entanglement attempts significantly impact scalability and overall system performance.

\subsection{Quantum Error Correction and Logical Qubits}
Fault-tolerant quantum computation relies on encoding quantum information into a block of $N$ physical bits as each logical qubit, which is intrinsically more robust against physical noise through redundancy and active error correction. The reliability of such encoded qubits is typically determined by code distance error $d$, which quantifies the error-correcting capability and directly impacts the scaling of the logical error rate. Under standard assumptions, the logical error rate $P_L$ per round of error correction can be approximated as Equation \ref{eq:logical_error}.
\begin{equation}
    p_L \approx C\left(\frac{p}{p_{th}} \right)^{(d+1)/2},
    \label{eq:logical_error}
\end{equation}
where $p$ is the physical error rate, $p_{th}$ is the error threshold of the QEC code\footnote{The maximum physical error rate below which QEC can successfully suppress errors by increasing the code distance.}, and $C$ is a constant that depends on the decoder and circuit-level implementation details \cite{PhysRevA.86.032324}. As indicated by (\ref{eq:logical_error}), increasing the code distance exponentially suppresses logical errors, at the cost of substantial overhead in physical qubits and operations.

Topological codes, particularly the surface code, have emerged as leading candidates for fault-tolerant architectures due to their locality constraints and relatively high error thresholds \cite{KITAEV20032, PhysRevA.83.020302, PhysRevA.86.032324}. However, logical operations in such codes must be carefully engineered to preserve fault tolerance, typically requiring structured sequences of physical gates, mid-circuit measurements (MCMs), classical feedforward, and continuous stabilization of the encoded state. These requirements introduce non-trivial overheads, especially for remote logical operations, where maintaining fault tolerance becomes increasingly challenging.

\subsection{Distributed Logical Gate Operations}
Realizing logical operations across spatially separated QPUs is a central challenge in DQC, as it requires coordinating encoded qubits under strict fault-tolerance constraints. A broad class of approaches relies on extending local fault-tolerant constructions to the distributed architectures, typically through entanglement-assisted designs. A straightforward strategy is to apply transversal operations across QPUs for Clifford gates. In this approach, a remote logical CNOT is decomposed into pairwise remote physical CNOT gates between corresponding qubits of the two code blocks, implemented via remote CNOT primitives as depicted in Fig.~\ref{fig:PhysRemoteCNOT}. This transversal-based design requires $N$ independent entanglement pairs, if logical qubits are encoded in $N$ physical qubits, resulting in a linear scaling of communication and entanglement resources with the code size \cite{Mueller2025-en}. While this method preserves fault tolerance due to its transversal structure, the associated entanglement overhead and synchronization cost become prohibitive for large code distances. An alternative class of methods realizes logical operations indirectly through parity measurements, which require an ancilla qubit to mediate joint measurements between data qubits. For instance, the CNOT gate can be performed as follows.
\begin{itemize}
    \item Perform a joint $Z \otimes Z$ measurement between the control qubit and ancilla qubit.
    \item Perform a joint $X \otimes X$ measurement between the ancilla qubit and the target qubit.
    \item Measure the ancilla qubit and apply conditional corrections to the control and target qubits.
\end{itemize}

This measurement-based construction has been extended to distributed settings using entanglement-assisted techniques \cite{Lee2023, PhysRevA.92.042305, Huai2024}. However, such approaches are typically designed at the physical level and do not directly account for the structure of encoded logical qubits. At the logical level, this method is typically realized through lattice surgery, particularly for topological codes such as the surface code. In this strategy, logical operations are implemented by dynamically merging and splitting code patches to perform joint stabilizer measurements \cite{Horsman_2012}.  While lattice surgery preserves locality and fault tolerance, it requires additional logical ancilla qubits and incurs non-trivial space-time overhead. In a distributed setting, it  also amplifies the dependence on high-fidelity entanglement and introduces additional communication latency and error sources in distributed architectures \cite{keskin2025lattice}. Indeed, for typical 2D surface codes whose code-distance $d$ (and $N$ is approximately $d^2$), a total of $O(d^2)$ Bell pairs will still be needed. This is because  in each round, $d$ Bell pairs are needed, and $O(d)$ rounds are needed for a remote CNOT based on lattice surgery \cite{remote-surface1, remote-surface2}.

\begin{figure}
    \centering
    \begin{subfigure}[b]{0.33\linewidth}
         \centering
         \includegraphics[width=\textwidth]{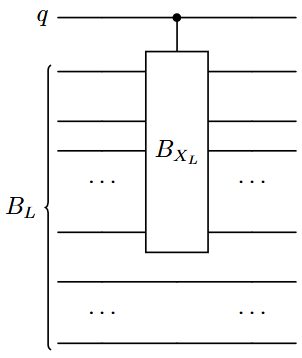}
         \caption{}
         \label{subfig:SingleControlDesired}
     \end{subfigure}
     \hfill
     \begin{subfigure}[b]{0.65\linewidth}
         \centering
         \includegraphics[width=\textwidth]{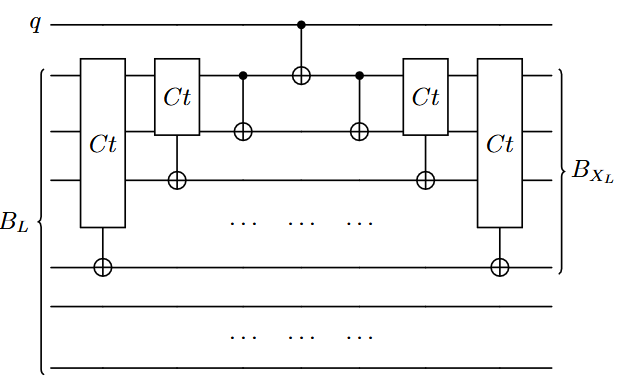}
         \caption{}
         \label{subfig:SingleControlNOBOL}
     \end{subfigure}
     \caption{CNOT circuit with single qubit as control. \textbf{(a)} The desired CNOT operation; $q$ is the control and $B_L$ is the target. \textbf{(b)} The proposed method; only qubits within $B_{X_L}$ need to be acted upon. The method is comprised entirely of CNOT gates. The target locations and a few controls are fixed, as shown in the figure, but most of the control qubits for each target have several possible locations. The $Ct$ rectangle denotes all possible locations for the control qubit, with an implicit requirement for symmetry in their locations in the circuit.}
     \label{fig:SingleQubitControl}
\end{figure}

Progress has been made towards more efficient logical operations. For instance, lattice surgery can reduce memory overhead by a factor of 4 \cite{fowler2019lowoverheadquantumcomputation}. This work also significantly reduces the overhead of magic state distillation and injection. More recent work has shown that noisy entanglement links can be tolerated in distributed lattice surgery when merging two patches \cite{Ramette2024}. This type of improvement would allow for less overhead with entanglement distillation. Another recent work presented a more general form of lattice surgery that can be applied to any quantum low density parity check code, and it reduces the number of stabilization required for the operation \cite{gj8x-n5gg}. This method could reduce the number of Bell pairs required in the distributed setting. Our work differs from these recent advances by reducing the number of required Bell pairs to only one, which is a significant improvement over the previously mentioned methods. NOBOL may require extra ancilla qubits, but the entire operation can be completed with a single entanglement connection.

\section{NOBOL: Proposed Design}

\begin{table}[b]
    \centering
    \caption{Symbols Used}
    \begin{tabular}{||c|c||}
    \hline
    \textbf{Symbol} & \textbf{Meaning} \\
    \hline
    $N_A$ & Total number of physical qubits in $A_L$ \\
    \hline
    $\tau_A$ & Number of qubits in the logical $Z$ operator for $A_L$ ($A_{Z_L}$) \\
    \hline
    $N_B$ & Total number of physical qubits in $B_L$ \\
    \hline 
    $\tau_B$ & Number of qubits in logical $X$ operator for $B_L$ ($B_{X_L}$) \\
    \hline
    \end{tabular}
    \label{tab:symbols}
\end{table}

In this section, we describe the design of NOBOL, a low-overhead, fault-tolerant logical CNOT operation between two remote logical qubits, $A_L$ (control) and $B_L$ (target), encoded using CSS codes, using only on Bell pair distributed at the same locations as $A_L$ and $B_L$, respectively, to minimize entanglement and ancilla overhead. To this end, we generalize the standard remote CNOT construction based on LOCC for physical qubits shown in Fig.~\ref{fig:PhysRemoteCNOT} and apply it at the logical level, while still leveraging only a single shared Bell pair, independent of the code size. Table~\ref{tab:symbols} shows the symbols we use throughout this section and their meanings. Any symbol without a subscript $A$ or $B$ is meant to refer to a general logical qubit, and not necessarily $A_L$ or $B_L$.

Below, we first develop efficient designs for implementing a local CNOT operation between a logical qubit and a single physical qubit (e.g., one half of the Bell pair), for the two cases where the former is the control and the latter is the target and vice versa.  We then show that combining the two designs is sufficient to construct a long-distance (or remote) logical CNOT between two encoded qubits, $A_L$ and $B_L$.

\subsection{Single Qubit as Control, Logical Qubit as Target}

The conventional physical-level remote CNOT gate requires a CNOT operation between one-half of a shared Bell pair and the target qubit. Extending this construction to the logical level necessitates the implementation of a CNOT gate in which a single physical qubit acts as the control and a logical qubit serves as the target. This operation constitutes a fundamental primitive underlying our proposed design of NOBOL.

More specifically, let $q$ denote a single control qubit in an arbitrary quantum state, and let $B_L$ represent a logical qubit encoded using a CSS code. By definition, the logical $X_L$ operator on $B_L$ can be implemented through applying a tensor product of Pauli-$X$ operations over a subset of its physical qubits, denoted by $B_{X_L}\subseteq B_L$. It should be noted that $B_{X_L}$ can be significantly smaller than the full code block $B_L$. For instance, in a surface code with $N$ physical qubits, the logical $X_L$ operator typically spans only $\sqrt{N}$ qubits. This observation leads to a fundamental intuition, in which the logical CNOT gate can be realized by interacting only with subset of the qubits, $B_{X_L}$, while leaving the remainder of the code block unaffected, as illustrated in Fig.~\ref{fig:SingleQubitControl}(a). Let $N_B=|B_L|$ and $\tau_B=|B_{X_L}|$ denote the size of the logical qubit $B$ and weight\footnote{The number of physical qubits on which $X_L$ acts non-trivially.} of logical $X_L$ operator on logical qubit $B$, respectively. The desired logical CNOT operation can then be expressed as the block-diagonal unitary matrix as Equation \ref{eq:logical_cnot_block}.
\begin{equation}
    U_{\text{L-CNOT}} =
    \begin{bmatrix}
        I^{\otimes \tau_B} & 0^{\otimes \tau_B} \\
        0^{\otimes \tau_B} & X^{\otimes \tau_B}
    \end{bmatrix},
    \label{eq:logical_cnot_block}
\end{equation}
which implements a logical CNOT by conditionally applying $X^{\otimes \tau_B}$ on the $B_{X_L}$ qubits. Since $X^{\otimes \tau_B}$ realizes the logical $X_L$ operation on the encoded qubit, this construction is equivalent to a CNOT between the control qubit and the logical qubit $B_L$. As illustrated in Fig.~\ref{fig:SingleQubitControl}(b), the proposed design admits $(\tau_B-1)!$ configurations with respect to the placement of control qubits, denoted by \emph{Ct} in the figure. It should be noted that symmetric configurations must be applied between the right and left sides of Fig.~\ref{fig:SingleQubitControl}(b). In all configurations, the CNOT operations are applied exclusively to the qubits in the support of the logical operator, $B_{X_L}$, while the remaining qubits in the code block $B_L$ remain unaffected. One of the fixed gates in the circuit is the CNOT operation between $q$ and first qubit in $B_{X_L}$. The other two fixed CNOT gates are a result of the placement rules for controls. Let the control qubit $q$ be assigned index $0$ and indices increase going downwards along the circuit. The placement of control operations must satisfy the following constraints:
\begin{enumerate}[label=\roman*.]
    \item The control for each CNOT gate must be placed on a qubit with a lower index than its corresponding target, excluding the qubit $q$ (i.e., index $0$).
    \item Each qubit (except index $1$) is leveraged as a target twice, and the corresponding control qubits must be placed on the same index (i.e., same qubit) to preserve the required conditional structure.
    \end{enumerate}

\subsection{Optimal-depth Construction of the Proposed Design}
\begin{figure}[b]
    \centering
    \includegraphics[width=\linewidth]{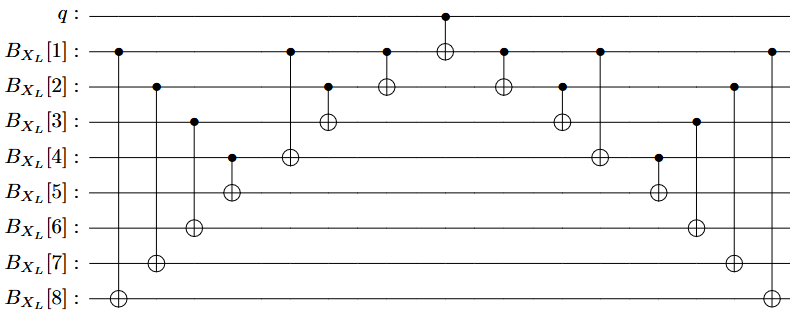}
    \caption{Depth Optimal CNOT for $|B_{X_L}|=8$}
    \label{fig:OptimalCNOT}
\end{figure}

As mentioned above, there are various configurations to realize the proposed structure by following the mentioned constraints. Circuit depth is a well-known evaluation metric, particularly in limited coherence circuits, to determine optimal construction. Therefore, in this section, we present an optimized construction that minimizes circuit depth while preserving the logical correctness of the operation. Algorithm~\ref{CNOTconstruction} presents a systematic procedure for synthesizing the circuit with logarithmic depth, achieving $O(\log \tau_B)$ scaling with respect to the weight of the logical $X_L$ operator on logical qubit $B_L$, $\tau_B = |B_{X_L}|$. This represents a significant improvement over naive constructions with linear depth. Fig.~\ref{fig:OptimalCNOT} illustrates an example of the resulting circuit for $|B_{X_L}| = 8$ qubits.

\begin{algorithm}
\caption{DepthOptimalCNOTConstruction}
\label{CNOTconstruction}
\begin{algorithmic}[1]
    \Input{Register $B_{X_L}$, Qubit $q$, Empty Circuit $C$}
    \Output{Circuit $C$ contains the gate operations to perform logical CNOT with $q$ as control and $B_L$ as the target}
    \State
    \LineComment{Base Case when $B_{X_L}$ has one qubit}
    \If{$|B_{X_L}| = 1$}
        \State append CNOT($q, B_{X_L}[1]$) to $C$
    \EndIf
    \State
    \LineComment{Recursive Case:}
    \LineComment{Compute the left side, apply recursion, compute the right side}
    \State $i\gets |B_{X_L}|, j\gets 1$
    \While{$i-j > 0$}
        \State append CNOT($j, i$) to $C$
        \State $i\gets i-1, j\gets j+1$
    \EndWhile
    \State
    \LineComment{$B_{X_L}[1:i]$ are the qubit indices from 1 to $i$. These qubits do not have targets yet.}
    \LineComment{$C[j:]$ represents the rest of circuit. There are no gate operations in this section yet.}
    \State DepthOptimalCNOTConstruction($B_{X_L}[1:i], q, C[j:])$
    \State
    \LineComment{Right side is a mirror image of the left side}
    \State $j\gets j-1$
    \While{$j>0$}
        \State append $C[j]$ to $C$
        \State $j\gets j-1$
    \EndWhile
    \Return{$C$}
    \LineComment{End of Algorithm DepthOptimalCNOTConstruction}
\end{algorithmic}
\end{algorithm}

\subsection{Logical Qubit as Control, Single Qubit as Target}

\begin{figure}
    \centering
    \includegraphics[width=0.8\linewidth]{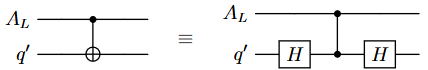}
    \caption{CNOT and CZ Equivalence}
    \label{fig:CNOTCZEquiv}
\end{figure}

\begin{figure}
    \centering
    \begin{subfigure}[b]{0.35\linewidth}
         \centering
         \includegraphics[width=\textwidth]{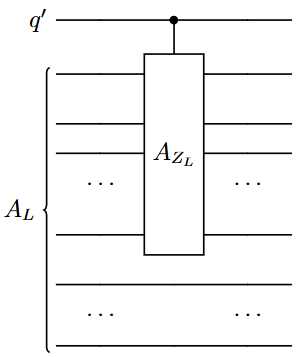}
         \caption{}
         \label{subfig:SingleCZDesired}
     \end{subfigure}
     \hfill
     \begin{subfigure}[b]{0.63\linewidth}
         \centering
         \includegraphics[width=\textwidth]{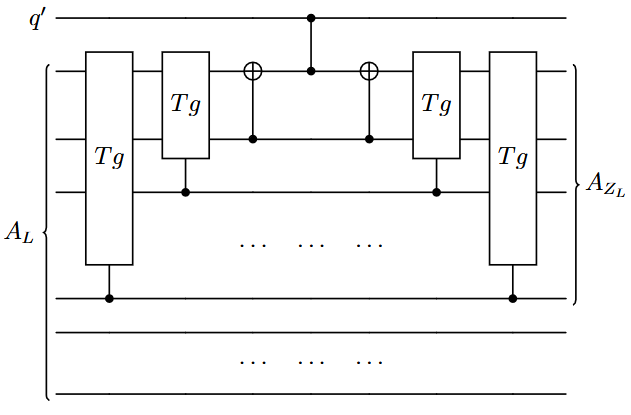}
         \caption{}
         \label{subfig:SingleCZNOBOL}
     \end{subfigure}
     \caption{CZ circuit with single qubit. \textbf{(a)} The desired CZ circuit between $q'$ and $A_L$. \textbf{(b)} The proposed CZ method; it is comprised of CNOT gates and a single CZ gate. The control locations for each CNOT gate are given. However, for most of the CNOT gates, there are several possible target locations for each control. The $Tg$ rectangles denote the possible qubit locations, with an implicit requirement for symmetry in their locations in the circuit.}
     \label{fig:SingleQubitCZ}
\end{figure}

In this section, we consider the realization of a CNOT operation where a logical qubit $A_L$ acts as the control and a single qubit $q'$ serves as the target.  Rather than implementing this operation directly, we exploit the well-known equivalence between CNOT and controlled-$Z$ (CZ) gates via Hadamard transformations, as illustrated in Fig.~\ref{fig:CNOTCZEquiv}. Specifically, the CNOT operation is equivalent to a CZ operation conjugated by Hadamard gates on the target qubit. Consequently, it suffices to construct a logical CZ operation between $A_L$ and $q'$, followed by a local Hadamard on $q'$, as illustrated in Fig.~\ref{fig:SingleQubitCZ}. 

Similar to the previous subsection, the key observation is that the logical action can be confined to the support of the logical CZ operator. In this case, as illustrated in Fig.~\ref{fig:SingleQubitCZ}(a), the operation only involves qubits in the support of the logical $Z_L$ operator, denoted by $A_{Z_L} \subseteq A_L$. Since $A_L$ is encoded using a CSS code, the logical $Z_L$ operator can be implemented as a tensor product of Pauli-$Z$ operators over this subset. Let $N_A = |A_L|$ and $\tau_A=|A_{Z_L}|$ denote the size of the code block B and weight of the logical $Z_L$ operator, respectively. The desired logical CZ operation can then be expressed as the block-diagonal unitary as Equation \ref{eq:logical_cz_block}.
\begin{equation}
    U_{\mathrm{L\text{-}CZ}}=
\begin{bmatrix}
    I^{\otimes \tau_A} & 0^{\otimes \tau_A} \\
    0^{\otimes \tau_A} & Z^{\otimes \tau_A}
    \end{bmatrix},
    \label{eq:logical_cz_block}
\end{equation}
which implements a logical CZ by conditionally applying $Z^{\otimes \tau_A}$ on the $A_{Z_L}$. Therefore, any quantum circuit that realizes Equation \ref{eq:logical_cz_block} constitutes a valid implementation of the logical CZ operation. 

Similar to the CNOT construction, the proposed design admits $(\tau_A-1)!$ distinct circuit realizations due to the flexibility in assigning control-target interactions. A general representation of construction is shown in Fig.~\ref{fig:SingleQubitCZ}(b), in which the proposed design admits multiple configurations with respect to the placement of control qubits, denoted by \emph{Tg} in the figure. One of the fixed gates in this circuit is the CZ gate between $q'$ and the first qubit of $A_{Z_L}$, serving as the anchor of the construction. The other fixed CNOT gates are a result of the rules for placing targets of each CNOT gate. The placement of target operations must satisfy the following constraints. Let the target qubit $q'$ be assigned index $0$, with indices increasing downward along the circuit.
\begin{enumerate}[label=\roman*.]
    \item The target of each CNOT gate must be placed on a qubit with a lower index than its corresponding control, excluding the qubit $q'$ (index $0$).
    \item Each qubit (except index $1$) is leveraged as a control twice, and the corresponding targets must be placed on the same index (i.e., same qubit) to preserve the required conditional structure.
\end{enumerate}

Referring back to Fig.~\ref{fig:CNOTCZEquiv}, we note that if we replace the only CZ gate in Fig.~\ref{fig:SingleQubitCZ}(b) with a CNOT gate with $q'$ as the target, the modified cricuit implements  CNOT (where $A_L$ is the control). This modified circuit thus has exactly $2\tau_A-1$ local CNOT gates.

\subsection{Optimal-depth CZ Construction}
\begin{figure}
    \centering
    \includegraphics[width=\linewidth]{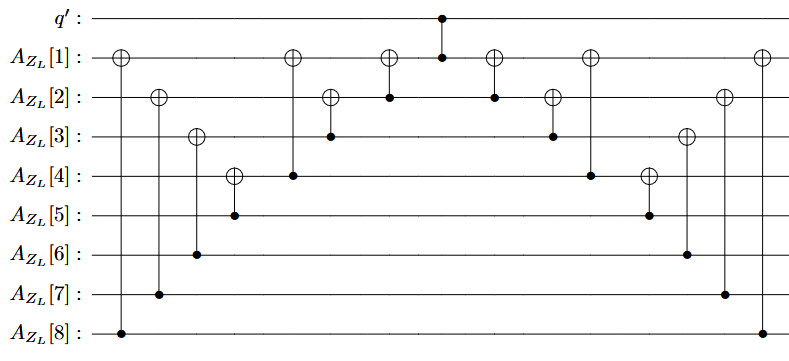}
    \caption{Depth Optimal CZ for $|A_{Z_L}|=8$}
    \label{fig:OptimalCZ}
\end{figure}

Due to the flexibility in assigning control–target pairs for the constituent CNOT gates, the proposed design admits $(\tau_A-1)!$ distinct circuit configurations, where $\tau_A = |A_{Z_L}|$ denotes the weight of the logical $Z_L$ operator on logical qubit $A_L$. To identify an optimal realization, we develop Algorithm~\ref{CZconstruction}, which provides a systematic procedure for constructing a circuit with minimal depth. In particular, the resulting construction achieves logarithmic depth scaling, i.e., $O(\log \tau_A)$, with respect to the size of the logical operator support. Fig.~\ref{fig:OptimalCZ} illustrates an example of the optimized CZ circuit for $|A_{Z_L}| = 8$ qubits. It is worth noting that the corresponding optimal circuit for CNOT where $A_L$ is the control would be identical except that the only CZ gate should be replaced with a CNOT gate where $q'$ is the target.

\begin{algorithm}
\caption{DepthOptimalCZConstruction}
\label{CZconstruction}
\begin{algorithmic}[1]
    \Input{Register $A_{Z_L}$, Qubit $q'$, Empty Circuit $C$}
    \Output{Circuit $C$ contains the gate operations to perform logical CZ between $q'$ and $A_L$}
    \State
    \LineComment{Base Case when $A_{Z_L}$ has one qubit}
    \If{$|A_{Z_L}| = 1$}
        \State append CZ($q', A_{Z_L}[1]$) to $C$
    \EndIf
    \State
    \LineComment{Recursive Case:}
    \LineComment{Compute the left side, apply recursion, compute the right side}
    \State $i\gets |A_{Z_L}|, j\gets 1$
    \While{$i-j > 0$}
        \State append CNOT($i, j$) to $C$
        \State $i\gets i-1, j\gets j+1$
    \EndWhile
    \State
    \LineComment{$A_{Z_L}[1:i]$ are the qubit indices from 1 to $i$. These qubits do not have controls yet.}
    \LineComment{$C[j:]$ represents the rest of circuit. There are no gate operations in this section yet.}
    \State OptimalConstructionCZ($A_{Z_L}[1:i], q', C[j:])$
    \State
    \LineComment{Right side is a mirror image of the left side}
    \State $j\gets j-1$
    \While{$j>0$}
        \State append $C[j]$ to $C$
        \State $j\gets j-1$
    \EndWhile
    \Return{$C$}
\end{algorithmic}
\end{algorithm}

\subsection{CNOT Between Two Logical Qubits}
We have introduced constructions for implementing CNOT operations between an arbitrary single physical qubit and an arbitrary logical qubit, covering both configurations in which the physical qubit acts as the control and as the target. Building on these primitives, we can introduce a logical CNOT operation between two encoded qubits. Let $A$ and $B$ in Fig.~\ref{fig:PhysRemoteCNOT} be replaced by logical qubits $A_L$ and $B_L$, respectively. The joint state of these logical qubits can be represented as an arbitrary two-logical-qubit state as follows:
\begin{equation}
\resizebox{0.89\linewidth}{!}{$
    \ket{\psi_L}=a\ket{00}_{A_LB_L} + b\ket{01}_{A_LB_L} + c\ket{10}_{A_LB_L} + d\ket{11}_{A_LB_L}$}
\end{equation}
By incorporating the shared Bell pair $\ket{\Phi^+}$, the joint state of the system can be expressed as:

\begin{equation}
\resizebox{0.91\linewidth}{!}{$
\begin{split}
    \ket{\psi}=\frac{1}{\sqrt{2}}(
    &a\ket{0}_{A_L} \ket{00}_{q'q} \ket{0}_{B_L} + a\ket{0}_{A_L} \ket{11}_{q'q} \ket{0}_{B_L} \\
    &+ b\ket{0}_{A_L} \ket{00}_{q'q} \ket{1}_{B_L} + b\ket{0}_{A_L} \ket{11}_{q'q} \ket{1}_{B_L} \\
    &+ c\ket{1}_{A_L} \ket{00}_{q'q} \ket{0}_{B_L} + c\ket{1}_{A_L} \ket{11}_{q'q} \ket{0}_{B_L} \\
    &+ d\ket{1}_{A_L} \ket{00}_{q'q} \ket{1}_{B_L} + d\ket{1}_{A_L} \ket{11}_{q'q} \ket{1}_{B_L}
    )
\end{split}$}
\end{equation}
After applying the CNOT operations between $A_L$, $B_L$, and the corresponding halves of the Bell pair, the resulting state can be expressed as:
\begin{equation}
\resizebox{0.91\linewidth}{!}{$
\begin{split}
    \ket{\psi}=\frac{1}{\sqrt{2}}(
    &a\ket{0}_{A_L} \ket{00}_{q'q} \ket{0}_{B_L} + a\ket{0}_{A_L} \ket{11}_{q'q} \ket{1}_{B_L} \\ 
    &+ b\ket{0}_{A_L} \ket{00}_{q'q} \ket{1}_{B_L} + b\ket{0}_{A_L} \ket{11}_{q'q} \ket{0}_{B_L} \\
    &+ c\ket{1}_{A_L} \ket{10}_{q'q} \ket{0}_{B_L} + c\ket{1}_{A_L} \ket{01}_{q'q} \ket{1}_{B_L} \\
    &+ d\ket{1}_{A_L} \ket{10}_{q'q} \ket{1}_{B_L} + d\ket{1}_{A_L} \ket{01}_{q'q} \ket{0}_{B_L}
    )
\end{split}
$}
\end{equation}
Applying a Hadamard gate to the second qubit of the Bell pair (qubit $q$) ensures the measurement is in the $X$-basis. The corresponding state immediately before measurement is expressed in (\ref{eq:BeforeMeasure}).
\begin{equation}
\begin{split}
    \ket{\psi}=\frac{1}{\sqrt{2}}(
    &\ket{0}_{A_L}  \ket{\psi_1} \ket{0}_{B_L} + \ket{0}_{A_L} \ket{\psi_2} \ket{1}_{B_L} \\
    &+ \ket{1}_{A_L} \ket{\psi_3} \ket{0}_{B_L} + \ket{1}_{A_L} \ket{\psi_4} \ket{1}_{B_L})
\end{split}
\label{eq:BeforeMeasure}
\end{equation}
where
\begin{equation}
\begin{split}
    \ket{\psi_1} &= \frac{1}{\sqrt{2}}(a(\ket{00}_{q'q} + \ket{01}_{q'q}) + b(\ket{10}_{q'q} - \ket{11}_{q'q})) \\
    \ket{\psi_2} &= \frac{1}{\sqrt{2}}(b(\ket{00}_{q'q} + \ket{01}_{q'q}) + a(\ket{10}_{q'q} - \ket{11}_{q'q})) \\
    \ket{\psi_3} &= \frac{1}{\sqrt{2}}(d(\ket{00}_{q'q} - \ket{01}_{q'q}) + c(\ket{10}_{q'q} + \ket{11}_{q'q})) \\
    \ket{\psi_4} &= \frac{1}{\sqrt{2}}(c(\ket{00}_{q'q} - \ket{01}_{q'q}) + d(\ket{10}_{q'q} + \ket{11}_{q'q})) \\
\end{split}
\end{equation}

At this stage, a $Z$-basis measurement is applied on $q'$ and $q$. The resulting measurement outcomes yield four possible combinations of classical bits, each corresponding to a distinct correction scenario. These outcomes determine the conditional logical operations that must be applied to $A_L$ and $B_L$. Table~\ref{Tab:Corrections} summarizes the required corrections and the corresponding resulting states. As shown, regardless of the measurement outcome, the final state is identical up to a global phase and can be expressed as follows:

\begin{equation}
\resizebox{0.89\linewidth}{!}{$
    \ket{\psi_L}=\ket{00}_{A_LB_L} + b\ket{01}_{A_LB_L} + c\ket{11}_{A_LB_L} + d\ket{10}_{A_LB_L}$}
\end{equation}

\begin{table*}
\caption{Measurement Result Corrections}
\begin{center}
\begin{tabular}{ || V{11em} || V{13em} ||  V{10em} || V{13em} || }
    \hline
    \textbf{Measurement Outcome} & \textbf{Normalized State} & \textbf{Correction Operation} & \textbf{Final State} \\
    \hline
    $00$ & $a\ket{00}_{A_LB_L}+b\ket{01}_{A_LB_L}$ $+c\ket{11}_{A_LB_L}+d\ket{10}_{A_LB_L}$ & None & $a\ket{00}_{A_LB_L}+b\ket{01}_{A_LB_L}$ $+c\ket{11}_{A_LB_L}+d\ket{10}_{A_LB_L}$ \\
    \hline
    $01$ & $a\ket{00}_{A_LB_L}+b\ket{01}_{A_LB_L}$ $-c\ket{11}_{A_LB_L}-d\ket{10}_{A_LB_L}$ & $Z_L\ket{A_L}$ & $a\ket{00}_{A_LB_L}+b\ket{01}_{A_LB_L}$ $+c\ket{11}_{A_LB_L}+d\ket{10}_{A_LB_L}$ \\
    \hline
    $10$ & $a\ket{01}_{A_LB_L}+b\ket{00}_{A_LB_L}$ $+c\ket{10}_{A_LB_L}+d\ket{11}_{A_LB_L}$ & $X_L\ket{B_L}$ & $a\ket{00}_{A_LB_L}+b\ket{01}_{A_LB_L}$ $+c\ket{11}_{A_LB_L}+d\ket{10}_{A_LB_L}$ \\
    \hline
    $11$ & $-a\ket{01}_{A_LB_L}-b\ket{00}_{A_LB_L}$ $+c\ket{10}_{A_LB_L}+d\ket{11}_{A_LB_L}$ & $Z_L\ket{A_L},X_L\ket{B_L}$ & $-a\ket{00}_{A_LB_L}-b\ket{01}_{A_LB_L}$ $-c\ket{11}_{A_LB_L}-d\ket{10}_{A_LB_L}$ \\
    \hline
\end{tabular}
\end{center}
\label{Tab:Corrections}
\end{table*}

\section{Fault Tolerance Enhancement of NOBOL}

In this section, we propose efficient error containment approaches that complement the proposed NOBOL design. We note that physical and logical qubit errors, as well as error propagation, occur in existing approaches are not unique to NOBOL. For example, in conventional transversal implementations, a physical qubit error of $A_L$ could propagate to $B_L$, potentially corrupting the encoded state. Therefore, our focus is not on eliminating errors,  but on mitigating the unique aspects of error propagation induced by the NOBOL construction.

More specifically, since NOBOL applied multiple CNOT operations on $A_{Z_L}$ and $B_{X_L}$, the errors are largely contained within those subsets of physical qubits. In addition, although a physical qubit error can shift its quantum state to any other quantum state,  the error can always be decomposed into a linear combination of Pauli-$X$ gates and Pauli-$Z$ gates. Consequently, it suffices to consider to fundamental error types, i.e., bit-flip errors ($X$-errors) and phase-flip errors ($Z$-errors). We first discuss the effect of such errors on the proposed design, and then introduce two novel approaches to mitigating the effect of error propagation, including using small repetition codes tailored to NOBOL and intermittent and adaptive stabilizers (as well as their combinations) while incurring minimal additional overhead.

\begin{figure}[h]
    \centering
    \includegraphics[width=0.8\linewidth]{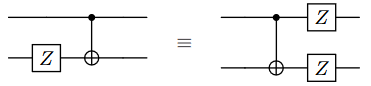}
    \caption{Phase-Flip Propagation in CNOT gate}
    \label{fig:SimplePhaseProp}
\end{figure}

\subsection{Phase-Flip Error Propagation}
\begin{figure}
    \centering
    \includegraphics[width=\linewidth]{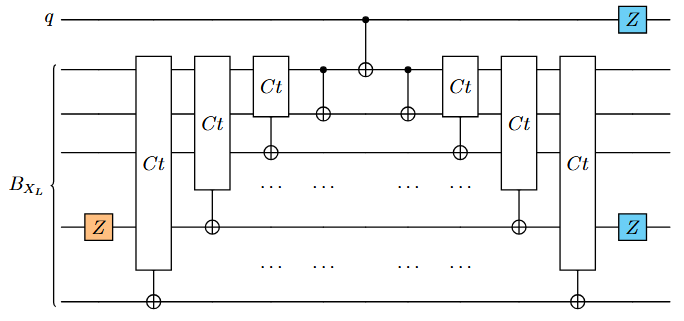}
    \caption{Phase Propagation with $B_{X_L}$. The orange $Z$ gate represents the initial error on qubit $i$ in the system. The blue $Z$ gates demonstrate how the error propagates. The result is an error on qubit $q$ and qubit $i$.}
    \label{fig:BXLPhaseProp}
\end{figure}

In a circuit with a single CNOT gate, a $Z$-error occurring on the target qubit prior to the operation propagates to the control qubit, as illustrated in Fig.~\ref{fig:SimplePhaseProp}. Since quantum circuits are reversible, simultaneous $Z$-errors on both the control and target qubit can be simplified to an equivalent $Z$-error on the target qubit. This behavior plays a key role in NOBOL, in particular, for the CNOT gate between the single qubit $q$ as the control and logical qubit $B_L$ as the target. The symmetrical layout of CNOT gates of the NOBOL constructions ensures that a single $Z$-error occurring within the $B_{X_L}$ does not propagate to other qubits in $B_L$, as shown in Fig.~\ref{fig:BXLPhaseProp}. However, such an error can propagate to the control qubit $q$, resulting in an incorrect measurement outcome. Consequently, the applied conditional correction on $A_L$ may be incorrect, leading to a logical $Z$-error on $A_L$.  Consequently, the applied conditional correction on $A_L$ may be incorrect, leading to a logical $Z$-error on $A_L$. This highlights a critical vulnerability of the construction and motivates the need for error containment strategies.

As for phase-flip error propagation in the CZ operation between $q'$ and $A_L$, the effect of the propagation depends on the layout of the CNOT gates. Under the optimal-depth construction, if a $Z$-error occurs on the $i$-th qubit of $A_{Z_L}$, the number of affected qubits after propagation is given by $|A_{Z_L}|/2^{i-1}$, assuming the first qubit of $A_{Z_L}$ is indexed as one. This relation indicates that errors originating at higher indices (i.e., lower qubit wires) propagate to fewer qubits.

\begin{figure}[b]
    \centering
    \includegraphics[width=0.8\linewidth]{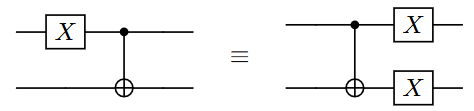}
    \caption{Bit-Flip Propagation in CNOT gate}
    \label{fig:SimpleBitProp}
\end{figure}

\subsection{Bit-Flip Error Propagation}
\begin{figure}
    \centering
    \includegraphics[width=\linewidth]{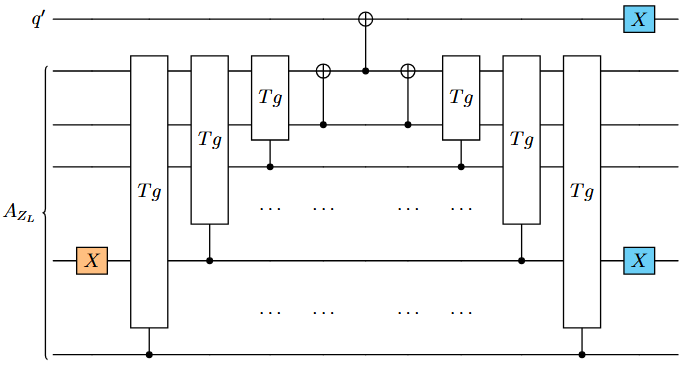}
    \caption{Bit Propagation with $A_{Z_L}$. The orange $X$ gate represents the initial error on qubit $i$ in the system. The blue $X$ gates demonstrate how the error propagates. The result is an error on qubit $q'$ and qubit $i$.}
    \label{fig:AZLBitProp}
\end{figure}

The CNOT gate also propagates $X$-error (bit-flip error), which exhibits complementary behavior to phase-flip errors. Specifically, for a single CNOT gate, an $X$-error acting on the control qubit prior to the operation propagates to the target qubit, as illustrated in Fig.~\ref{fig:SimpleBitProp}. In the NOBOL design, this propagation behavior is similarly constrained by the circuit structure.  In particular, for the CZ-based interaction between $q'$ and $A_L$, the symmetric layout of the construction induces partial cancellation of propagated errors, ensuring that a single $X$-error remains localized and affects the qubit $q'$. However, such an error can still lead to an incorrect measurement outcome, which in turn results in an erroneous conditional correction and induces a logical bit-flip error on $B_L$. Fig.~\ref{fig:AZLBitProp} illustrates this propagation behavior in the general construction.

As with phase-flip error propagation in $A_{Z_L}$, bit-flip error propagation within $B_{X_L}$ is confined to the second half of the construction. Similarly, the effect of bit-flip error propagation depends on the layout of the underlying CNOT gates. Under the optimal-depth construction, if an $X$-error occurs on the $i$-th qubit of $B_{X_L}$, the number of affected qubits after propagation is given by $|B_{X_L}|/2^{i-1}$. This relation indicates that errors originating at higher indices propagate to fewer qubits.

\subsection{Error Containment Techniques for NOBOL}
The previous section demonstrated how error propagation can affect the correctness of the proposed operations. In this section, we introduce two complementary error containment strategies tailored to the NOBOL design. These techniques can be applied independently or jointly to effectively suppress error propagation while maintaining low overhead.

\subsubsection{Repetition Codes}
Repetition codes provide a simple yet effective mechanism for introducing redundancy by encoding a single qubit across multiple physical qubits. For instance, a bit-flip repetition code maps
\begin{equation}
    \alpha\ket{0}+\beta\ket{1} \;\rightarrow\; \alpha\ket{0}^{\otimes n_r}+\beta\ket{1}^{\otimes n_r},
\end{equation}
where $n_r$ denotes the repetition factor, while a phase repetition code similarly encodes states in the $\{\ket{+},\ket{-}\}$ basis.

\begin{figure*}[t]
    \centering
    \captionsetup[subfigure]{oneside,margin={0.7cm,0cm}}
    \begin{subfigure}[b]{0.32\linewidth}
        \centering
        \includegraphics[width=\linewidth]{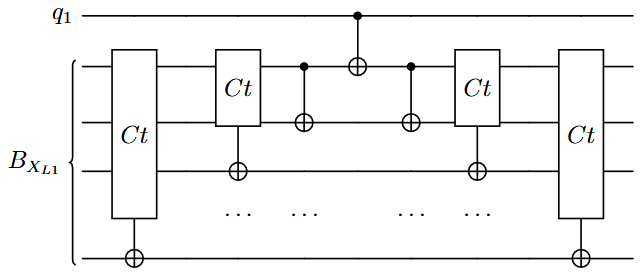}
        \caption{}
        \label{subfig:BitRep1}
    \end{subfigure}
    \hfill
    \begin{subfigure}[b]{0.32\linewidth}
        \centering
        \includegraphics[width=\linewidth]{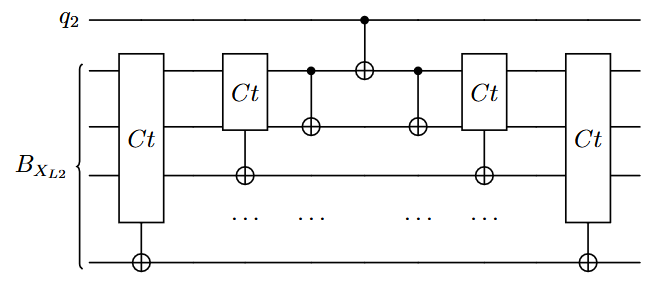}
        \caption{}
        \label{subfig:BitRep1}
    \end{subfigure}
    \hfill
    \begin{subfigure}[b]{0.32\linewidth}
        \centering
        \includegraphics[width=\linewidth]{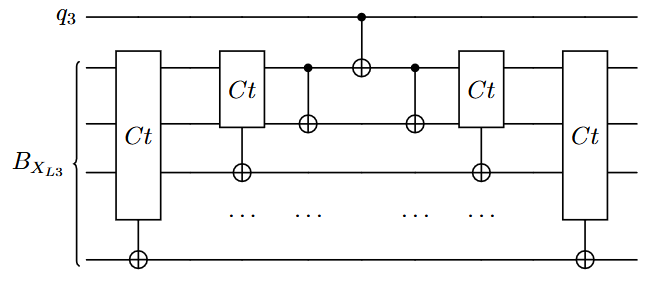}
        \caption{}
        \label{subfig:BitRep1}
    \end{subfigure}
    \caption{$q_1$, $q_2$, and $q_3$ form a three qubit bit repetition code. Each qubit $q_i$ independently performs a CNOT operation with each chunk $B_{X_{Li}}$ which, all together, forms a CNOT operation between a three qubit repetition code and $B_L$.}
    \label{fig:BitRepetitionCode}
\end{figure*}

In the NOBOL design, in particular for the CNOT between $A_L$ and $B_L$, the shared Bell pair can be encoded using small repetition codes to enhance robustness. Specifically, qubit $q$ is encoded using a bit-flip repetition code, while $q'$ is encoded using a phase-flip repetition code to enhance their resilience to error propagation. For the CNOT operation between $q$ and $B_L$, if $q$ is encoded using a repetition code of size $n_r$, where $3 \leq n_r \leq |B_{X_L}|$, the $B_{X_L}$ can be partitioned into $n_r$ chunks (or sub-blocks). Each encoded qubit $q_i$ then independently interacts with a corresponding subset $B_{X_{L_i}}$. The partitioning of the circuit into $n_r=3$ sub-blocks is illustrated in Fig.~\ref{fig:BitRepetitionCode}. The proposed solution provides two key advantages. 
\vspace{-0.2em}
\begin{enumerate}[label=\roman*.]
    \item \textbf{Shallow-depth Circuits:} Independent interactions enable simultaneous execution of multiple CNOT operations, which increases parallelism, while reducing circuit depth, which eliminates decoherence concerns.
    \item \textbf{Error Containment:} Bit-flip errors remain localized within their respective sub-blocks, increasing the tolerance of the NOBOL to bit-flip errors.
\end{enumerate}
\vspace{-0.3em}

However, this approach provides asymmetric protection, in which bit-flip repetition codes protect against $X$-errors but not $Z$-errors, while phase-flip repetition codes exhibit the opposite behavior. Consequently, a trade-off arises such that increasing the repetition size $n_r$ improves error tolerance at the cost of additional ancilla resources.

\begin{figure*}[t]
    \centering
    \begin{subfigure}[b]{0.32\linewidth}
        \centering
        \includegraphics[width=\linewidth]{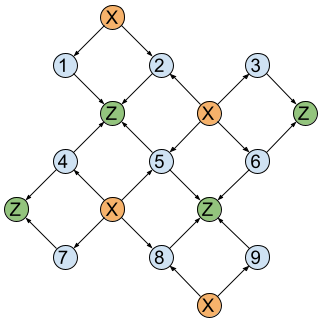}
        \caption{Original $3 \times 3$ surface code as $B_L$.}
        \label{subfig:SurfaceCode}
    \end{subfigure}
    \hfill
    \begin{subfigure}[b]{0.32\linewidth}
        \centering
        \includegraphics[width=\linewidth]{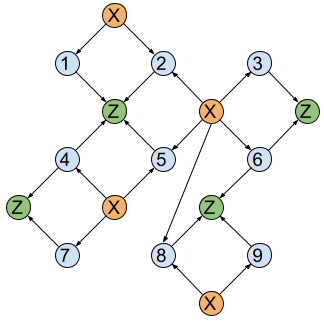}
        \caption{After CNOT(5, 8).}
        \label{subfig:SurfaceCodeCNOT1}
    \end{subfigure}
    \hfill
    \begin{subfigure}[b]{0.32\linewidth}
        \centering
        \includegraphics[width=\linewidth]{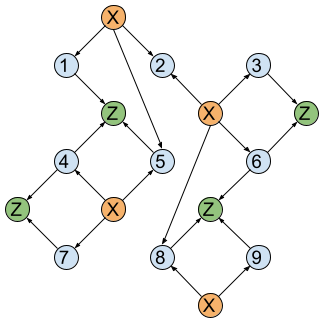}
        \caption{After CNOT(2, 5).}
        \label{subfig:SurfaceCodeCNOT2}
    \end{subfigure}
    \caption{Evolution of surface code stabilizers during the CNOT circuit. Blue qubits represent data qubits, green qubits correspond to $Z$ stabilizers, and orange qubits denote $X$ stabilizers. Arrows indicate the direction of CNOT gates used for stabilization.}
    \label{fig:SurfaceCodeStabilizers}
\end{figure*}

\subsubsection{Intermittent and Adaptive Stabilization}
A complementary approach to mitigating error propagation is to perform stabilization intermittently throughout the circuit execution. This strategy suppresses error accumulation by correcting errors before they propagate across multiple qubits. A key feature of NOBOL is that, although stabilizers evolve during the execution of intermediate CNOT gates, their evolution follows deterministic transformation rules. For a CNOT gate with control qubit $c$ and target qubit $t$, the stabilizer transformations are given by Equation \ref{eq:stabilizer-transformation}.
\begin{equation}
\label{eq:stabilizer-transformation}
\begin{aligned}
    X_c &\rightarrow X_c X_t, \quad & X_t &\rightarrow X_t \\
    Z_c &\rightarrow Z_c, \quad & Z_t &\rightarrow Z_c Z_t
\end{aligned}
\end{equation}
\vspace{-1.3em}

These predictable transformations enable adaptive stabilization, where stabilizers are updated dynamically to remain consistent with the evolving circuit. To illustrate the proposed stabilization strategy, consider the logical qubit $B_L$ encoded using a $3 \times 3$ surface code, as illustrated in Fig.~\ref{fig:SurfaceCodeStabilizers}(a), which a CNOT operation is performed between $q$ and $B_L$. In this case, one valid choice of the logical operator support is $B_{X_L}=\{2, 5, 8\}$. Accordingly, a valid sequence of CNOT operations is given by $\mathrm{CNOT}(5,8)$, $\mathrm{CNOT}(2,5)$, $\mathrm{CNOT}(q,2)$, $\mathrm{CNOT}(2,5)$, $\mathrm{CNOT}(5,8)$. Fig.~\ref{fig:SurfaceCodeStabilizers} illustrates he evolution of stabilizers throughout this sequence. After $\mathrm{CNOT}(5,8)$, the stabilizers are updated as Equation \ref{eq:firt-gate}.
\begin{equation}
\label{eq:firt-gate}
\begin{split}
    X_2X_3X_5X_6 &\rightarrow X_2X_3X_5X_6X_8 \\
    X_4X_5X_7X_8 &\rightarrow X_4X_5X_7 \\
    Z_5Z_6Z_8Z_9 &\rightarrow Z_6Z_8Z_9
\end{split}
\end{equation}
This transformation corresponds to Fig.~\ref{fig:SurfaceCodeStabilizers}(b). As shown in Fig.~\ref{fig:SurfaceCodeStabilizers}(c), after $\mathrm{CNOT}(2,5)$, the stabilizers evolve as Equation \ref{eq:second-gate}.

\begin{equation}
\label{eq:second-gate}
\begin{split}
    X_1X_2 &\rightarrow X_1X_2X_5 \\
    X_2X_3X_5X_6X_8 &\rightarrow X_2X_3X_6X_8 \\
    Z_1Z_2Z_4Z_5 &\rightarrow Z_1Z_4Z_5
\end{split}
\end{equation}

The operation $\mathrm{CNOT}(q,2)$ does not modify the stabilizers of $B_L$, as it does not involve an internal CNOT gate. The subsequent operations reverse the earlier transformations, such that after the second $\mathrm{CNOT}(2,5)$, the stabilizers revert to those in Fig.~\ref{fig:SurfaceCodeStabilizers}(b), and after the final $\mathrm{CNOT}(5,8)$, the stabilizers return to their original configuration in Fig.~\ref{fig:SurfaceCodeStabilizers}(a).

This example demonstrates that stabilizers can be updated consistently throughout the circuit, enabling intermediate error correction without disrupting the logical state. However, the frequency of stabilization introduces a trade-off between error suppression and execution time. While performing stabilization after each CNOT gate minimizes error propagation, it increases circuit depth. In general, each stabilization round incurs $O(d)$ depth, where $d$ is the code distance. Since the circuit consists of $2\tau_B$ CNOT operations, where $\tau_B = |B_{X_L}|$, the worst-case depth scales as $2\tau_B*O(d)=O(\tau_Bd)$. A similar analysis applies to the CNOT operation between $q'$ and $A_L$.

\subsubsection{Combined Error Containment Strategy}
The above techniques can be combined to further enhance fault tolerance. Specifically, repetition codes localize errors, while intermittent stabilization removes accumulated errors across the entire code. This combined strategy enables simultaneous containment of both bit-flip and phase-flip errors.

While the combined approach improves robustness, it introduces a trade-off between resource overhead and execution latency. Larger repetition codes require additional ancilla qubits, while more frequent stabilization increases circuit depth. Nevertheless, these trade-offs can be tuned based on hardware constraints and target logical error rates, making the approach highly adaptable to practical quantum computing systems.

\section{Conclusion}
In this paper, we have introduced NOBOL, a novel low-overhead approach for realizing \textit{long-distance or non-neighboring, logical gate} operations using only a single entangled state, along with local operations and classical computing. It enables efficient implementation of logical CNOT operations across both monolithic and distributed quantum architectures.

We have also illustrated the structural flexibility of the proposed design, which admits a factorial number of circuit configurations due to the freedom in qubit ordering that makes it well-suited across various qubit modalities and connectivity constraints. We have demonstrated that NOBOL admits a depth-optimal construction with logarithmic depth scaling, making it well-aligned with coherence-time limitations in fault-tolerant quantum systems. Moreover, the code-agnostic nature of the NOBOL allows it to be applied to a broad class of CSS codes, including the well-known surface codes. Furthermore, we developed unique containment techniques for error propagation, including intermittent and adaptive stabilization and encoding each half of the entangled pair using small repetition codes, which provide a flexible trade-off between ancilla qubits and execution time. 

In summary, NOBOL offers a practical and scalable solution for large-scale fault-tolerant quantum computing, particularly in architectures with limited inter- and intra-QPU connectivity and constrained ancilla resources. It also greatly benefits logical gate operation across a Quantum Internet where networking resources are even more scarce and end-to-end entangled connections are even more fragile. While this work has laid down the solid theoretical and algorithmic foundations, there remains much further work, including quantitative evaluation of fault-tolerance performance, and the design and analysis of optimal error-containment strategies.

\bibliographystyle{IEEEtran}
\bibliography{references}

\begin{thebibliography}{10}
\providecommand{\url}[1]{#1}
\csname url@samestyle\endcsname
\providecommand{\newblock}{\relax}
\providecommand{\bibinfo}[2]{#2}
\providecommand{\BIBentrySTDinterwordspacing}{\spaceskip=0pt\relax}
\providecommand{\BIBentryALTinterwordstretchfactor}{4}
\providecommand{\BIBentryALTinterwordspacing}{\spaceskip=\fontdimen2\font plus
\BIBentryALTinterwordstretchfactor\fontdimen3\font minus \fontdimen4\font\relax}
\providecommand{\BIBforeignlanguage}[2]{{%
\expandafter\ifx\csname l@#1\endcsname\relax
\typeout{** WARNING: IEEEtran.bst: No hyphenation pattern has been}%
\typeout{** loaded for the language `#1'. Using the pattern for}%
\typeout{** the default language instead.}%
\else
\language=\csname l@#1\endcsname
\fi
#2}}
\providecommand{\BIBdecl}{\relax}
\BIBdecl

\bibitem{doi:10.1137/S0097539795293172}
\BIBentryALTinterwordspacing
P.~W. Shor, ``Polynomial-time algorithms for prime factorization and discrete logarithms on a quantum computer,'' \emph{SIAM Journal on Computing}, vol.~26, no.~5, pp. 1484--1509, 1997. [Online]. Available: \url{https://doi.org/10.1137/S0097539795293172}
\BIBentrySTDinterwordspacing

\bibitem{10.1145/3708471}
\BIBentryALTinterwordspacing
O.~Regev, ``An efficient quantum factoring algorithm,'' \emph{J. ACM}, vol.~72, no.~1, Jan. 2025. [Online]. Available: \url{https://doi.org/10.1145/3708471}
\BIBentrySTDinterwordspacing

\bibitem{Cerezo2022}
\BIBentryALTinterwordspacing
M.~Cerezo, K.~Sharma, A.~Arrasmith, and P.~J. Coles, ``Variational quantum state eigensolver,'' \emph{npj Quantum Information}, vol.~8, no.~1, p. 113, Sep 2022. [Online]. Available: \url{https://doi.org/10.1038/s41534-022-00611-6}
\BIBentrySTDinterwordspacing

\bibitem{128057}
R.~P. Feynman, ``There's plenty of room at the bottom [data storage],'' \emph{Journal of Microelectromechanical Systems}, vol.~1, no.~1, pp. 60--66, 1992.

\bibitem{babaie2024}
S.~Babaie and C.~Qiao, ``Towards distributed quantum error correction for distributed quantum computing,'' in \emph{IEEE International Conference on Quantum Communications, Networking, and Computing (QCNC)}, 2025, pp. 66--73.

\bibitem{Campbell2024}
\BIBentryALTinterwordspacing
E.~Campbell, ``A series of fast-paced advances in quantum error correction,'' \emph{Nature Reviews Physics}, vol.~6, no.~3, pp. 160--161, Mar 2024. [Online]. Available: \url{https://doi.org/10.1038/s42254-024-00706-3}
\BIBentrySTDinterwordspacing

\bibitem{qiao-qdn-dqc}
C.~Qiao, Y.~Zhao, G.~Zhao, and H.~Xu, ``Quantum data networking for distributed quantum computing: Opportunities and challenges,'' in \emph{IEEE INFOCOM 2022-IEEE Conference on Computer Communications Workshops (INFOCOM WKSHPS)}.\hskip 1em plus 0.5em minus 0.4em\relax IEEE, 2022, pp. 1--6.

\bibitem{zhao-qdn}
Y.~Zhao and C.~Qiao, ``Distributed transport protocols for quantum data networks,'' \emph{IEEE/ACM Transactions on Networking}, vol.~31, no.~6, pp. 2777--2792, 2023.

\bibitem{filippov2025architectingdistributedquantumcomputers}
\BIBentryALTinterwordspacing
D.~Filippov, P.~Yang, and P.~Murali, ``Architecting distributed quantum computers: Design insights from resource estimation,'' 2025. [Online]. Available: \url{https://arxiv.org/abs/2508.19160}
\BIBentrySTDinterwordspacing

\bibitem{mohseni2026buildquantumsupercomputerscaling}
\BIBentryALTinterwordspacing
M.~Mohseni, A.~Scherer, K.~G. Johnson \emph{et~al.}, ``How to build a quantum supercomputer: Scaling from hundreds to millions of qubits,'' 2026. [Online]. Available: \url{https://arxiv.org/abs/2411.10406}
\BIBentrySTDinterwordspacing

\bibitem{9923784}
K.~N. Smith, G.~S. Ravi, J.~M. Baker, and F.~T. Chong, ``Scaling superconducting quantum computers with chiplet architectures,'' in \emph{2022 55th IEEE/ACM International Symposium on Microarchitecture (MICRO)}, 2022, pp. 1092--1109.

\bibitem{VanDamme2024}
\BIBentryALTinterwordspacing
J.~Van~Damme, S.~Massar, R.~Acharya, T.~Ivanov, D.~Perez~Lozano, Y.~Canvel, M.~Demarets, D.~Vangoidsenhoven, Y.~Hermans, J.~G. Lai, A.~M. Vadiraj, M.~Mongillo, D.~Wan, J.~De~Boeck, A.~Poto{\v{c}}nik, and K.~De~Greve, ``Advanced cmos manufacturing of superconducting qubits on 300 mm wafers,'' \emph{Nature}, vol. 634, no. 8032, pp. 74--79, Oct 2024. [Online]. Available: \url{https://doi.org/10.1038/s41586-024-07941-9}
\BIBentrySTDinterwordspacing

\bibitem{Hertzberg2021}
\BIBentryALTinterwordspacing
J.~B. Hertzberg, E.~J. Zhang, S.~Rosenblatt, E.~Magesan, J.~A. Smolin, J.-B. Yau, V.~P. Adiga, M.~Sandberg, M.~Brink, J.~M. Chow, and J.~S. Orcutt, ``Laser-annealing josephson junctions for yielding scaled-up superconducting quantum processors,'' \emph{npj Quantum Information}, vol.~7, no.~1, p. 129, Aug 2021. [Online]. Available: \url{https://doi.org/10.1038/s41534-021-00464-5}
\BIBentrySTDinterwordspacing

\bibitem{babaie2025}
S.~Babaie and C.~Qiao, ``Distributed quantum bit-phase-flip error correction based on css codes and ghz states,'' in \emph{45th IEEE International Conference on Distributed Computing Systems Workshops (ICDCSW)}, 2025, pp. 195--200.

\bibitem{11568217}
S.~Babaie, S.~Grzenda, and C.~Qiao, ``Distributed quantum error correction: Advancements and future research directions,'' in \emph{2026 24th International Symposium on Modeling and Optimization in Mobile, Ad Hoc, and Wireless Networks (WiOpt)}, 2026, pp. 1--8.

\bibitem{Mueller2025-en}
F.~Mueller, M.~Wang, and J.~Stack, ``Transversal fault tolerant distributed quantum computing operations,'' Sep. 2025.

\bibitem{PhysRevA.92.042305}
\BIBentryALTinterwordspacing
J.~Govenius, Y.~Matsuzaki, I.~G. Savenko, and M.~M\"ott\"onen, ``Parity measurement of remote qubits using dispersive coupling and photodetection,'' \emph{Phys. Rev. A}, vol.~92, p. 042305, Oct 2015. [Online]. Available: \url{https://link.aps.org/doi/10.1103/PhysRevA.92.042305}
\BIBentrySTDinterwordspacing

\bibitem{Huai2024}
\BIBentryALTinterwordspacing
S.~Huai, K.~Bu, X.~Gu, Z.~Zhang, S.~An, X.~Yang, Y.~Li, T.~Cai, and Y.~Zheng, ``Fast joint parity measurement via collective interactions induced by stimulated emission,'' \emph{Nature Communications}, vol.~15, no.~1, p. 3045, Apr 2024. [Online]. Available: \url{https://doi.org/10.1038/s41467-024-47379-1}
\BIBentrySTDinterwordspacing

\bibitem{keskin2025lattice}
B.~Keskin, C.~Afradi, S.~Lovis, M.~Palesi, P.~Escofet, C.~G. Almudever, and E.~Charbon, ``Lattice surgery aware resource analysis for the mapping and scheduling of quantum circuits for scalable modular architectures,'' \emph{arXiv preprint arXiv:2511.21885v1}, 2025.

\bibitem{PhysRevA.62.052317}
\BIBentryALTinterwordspacing
J.~Eisert, K.~Jacobs, P.~Papadopoulos, and M.~B. Plenio, ``Optimal local implementation of nonlocal quantum gates,'' \emph{Phys. Rev. A}, vol.~62, p. 052317, Oct 2000. [Online]. Available: \url{https://link.aps.org/doi/10.1103/PhysRevA.62.052317}
\BIBentrySTDinterwordspacing

\bibitem{11000201}
C.~Zhan, J.~Chung, A.~Zang, A.~Kolar, and R.~Kettimuthu, ``Design and simulation of the adaptive continuous entanglement generation protocol,'' in \emph{2025 International Conference on Quantum Communications, Networking, and Computing (QCNC)}, 2025, pp. 127--134.

\bibitem{PhysRevA.107.022428}
\BIBentryALTinterwordspacing
H.~Zhou, T.~Li, and K.~Xia, ``Parallel and heralded multiqubit entanglement generation for quantum networks,'' \emph{Phys. Rev. A}, vol. 107, p. 022428, Feb 2023. [Online]. Available: \url{https://link.aps.org/doi/10.1103/PhysRevA.107.022428}
\BIBentrySTDinterwordspacing

\bibitem{doi:10.1126/science.aam9288}
\BIBentryALTinterwordspacing
S.~Wehner, D.~Elkouss, and R.~Hanson, ``Quantum internet: A vision for the road ahead,'' \emph{Science}, vol. 362, no. 6412, p. eaam9288, 2018. [Online]. Available: \url{https://www.science.org/doi/abs/10.1126/science.aam9288}
\BIBentrySTDinterwordspacing

\bibitem{RevModPhys.83.33}
\BIBentryALTinterwordspacing
N.~Sangouard, C.~Simon, H.~de~Riedmatten, and N.~Gisin, ``Quantum repeaters based on atomic ensembles and linear optics,'' \emph{Rev. Mod. Phys.}, vol.~83, pp. 33--80, Mar 2011. [Online]. Available: \url{https://link.aps.org/doi/10.1103/RevModPhys.83.33}
\BIBentrySTDinterwordspacing

\bibitem{doi:10.1126/science.aao4309}
\BIBentryALTinterwordspacing
C.~Neill, P.~Roushan, K.~Kechedzhi, S.~Boixo, S.~V. Isakov, V.~Smelyanskiy, A.~Megrant, B.~Chiaro, A.~Dunsworth, K.~Arya, R.~Barends, B.~Burkett, Y.~Chen, Z.~Chen, A.~Fowler, B.~Foxen, M.~Giustina, R.~Graff, E.~Jeffrey, T.~Huang, J.~Kelly, P.~Klimov, E.~Lucero, J.~Mutus, M.~Neeley, C.~Quintana, D.~Sank, A.~Vainsencher, J.~Wenner, T.~C. White, H.~Neven, and J.~M. Martinis, ``A blueprint for demonstrating quantum supremacy with superconducting qubits,'' \emph{Science}, vol. 360, no. 6385, pp. 195--199, 2018. [Online]. Available: \url{https://www.science.org/doi/abs/10.1126/science.aao4309}
\BIBentrySTDinterwordspacing

\bibitem{PhysRevA.86.032324}
\BIBentryALTinterwordspacing
A.~G. Fowler, M.~Mariantoni, J.~M. Martinis, and A.~N. Cleland, ``Surface codes: Towards practical large-scale quantum computation,'' \emph{Phys. Rev. A}, vol.~86, p. 032324, Sep 2012. [Online]. Available: \url{https://link.aps.org/doi/10.1103/PhysRevA.86.032324}
\BIBentrySTDinterwordspacing

\bibitem{KITAEV20032}
\BIBentryALTinterwordspacing
A.~Kitaev, ``Fault-tolerant quantum computation by anyons,'' \emph{Annals of Physics}, vol. 303, no.~1, pp. 2--30, 2003. [Online]. Available: \url{https://www.sciencedirect.com/science/article/pii/S0003491602000180}
\BIBentrySTDinterwordspacing

\bibitem{PhysRevA.83.020302}
\BIBentryALTinterwordspacing
D.~S. Wang, A.~G. Fowler, and L.~C.~L. Hollenberg, ``Surface code quantum computing with error rates over 1
\BIBentrySTDinterwordspacing

\bibitem{Lee2023}
\BIBentryALTinterwordspacing
S.-H. Lee, S.~Omkar, Y.~S. Teo, and H.~Jeong, ``Parity-encoding-based quantum computing with bayesian error tracking,'' \emph{npj Quantum Information}, vol.~9, no.~1, p.~39, Apr 2023. [Online]. Available: \url{https://doi.org/10.1038/s41534-023-00705-9}
\BIBentrySTDinterwordspacing

\bibitem{Horsman_2012}
\BIBentryALTinterwordspacing
D.~Horsman, A.~G. Fowler, S.~Devitt, and R.~V. Meter, ``Surface code quantum computing by lattice surgery,'' \emph{New Journal of Physics}, vol.~14, no.~12, p. 123011, dec 2012. [Online]. Available: \url{https://doi.org/10.1088/1367-2630/14/12/123011}
\BIBentrySTDinterwordspacing

\bibitem{remote-surface1}
T.~H. Haug, T.~Hillmann, A.~F. Kockum, and R.~Van~Laer, ``Lattice surgery with bell measurements: Modular fault-tolerant quantum computation at low entanglement cost,'' \emph{arXiv preprint arXiv:2510.13541}, 2025.

\bibitem{remote-surface2}
S.~Liu, J.~Stack, K.~Sun, R.~Van~Beeumen, I.~Monga, K.~Klymko, K.~R. Brown, and E.~Saglamyurek, ``Remote entanglement in lattice surgery: To distill, or not to distill,'' \emph{arXiv preprint arXiv:2603.06513}, 2026.

\bibitem{fowler2019lowoverheadquantumcomputation}
\BIBentryALTinterwordspacing
A.~G. Fowler and C.~Gidney, ``Low overhead quantum computation using lattice surgery,'' 2019. [Online]. Available: \url{https://arxiv.org/abs/1808.06709}
\BIBentrySTDinterwordspacing

\bibitem{Ramette2024}
\BIBentryALTinterwordspacing
J.~Ramette, J.~Sinclair, N.~P. Breuckmann, and V.~Vuleti{\'{c}}, ``Fault-tolerant connection of error-corrected qubits with noisy links,'' \emph{npj Quantum Information}, vol.~10, no.~1, p.~58, Jun 2024. [Online]. Available: \url{https://doi.org/10.1038/s41534-024-00855-4}
\BIBentrySTDinterwordspacing

\bibitem{gj8x-n5gg}
\BIBentryALTinterwordspacing
A.~Cowtan, Z.~He, D.~J. Williamson, and T.~J. Yoder, ``Parallel logical measurements via quantum code surgery,'' \emph{PRX Quantum}, vol.~7, p. 020325, May 2026. [Online]. Available: \url{https://link.aps.org/doi/10.1103/gj8x-n5gg}
\BIBentrySTDinterwordspacing

\end{thebibliography}

\end{document}